# Journal of Materials Chemistry C

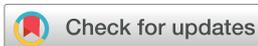

# Efficient generation of highly crystalline carbon quantum dots *via* electrooxidation of ethanol for rapid photodegradation of organic dyes†

Santiago D. Barrionuevo, [a] Federico Fioravanti,[b] Jorge M. Nuñez,[cdefg] Mauricio Llaver,[h] Myriam H. Aguirre, [fgi] Martin G. Bellino, [j] Gabriela I. Lacconi[b] and Francisco J. Ibañez [*a]

Achieving versatile routes to generate crystalline carbon-based nanostructures has become a fervent pursuit in photocatalysis-related fields. We demonstrate that the direct electrooxidation of ethanol, performed on Ni foam, yields ultra-small and highly crystalline graphene-like structures named carbon quantum dots (CQDs). We perform simulations of various $sp^2$ and $sp^3$ domains in order to understand the optical properties of CQDs by accounting their contribution as absorbance/luminescent centers in the overall optical response. Experiments and simulations reveal that absorbance bands for as-synthesized CQDs are dominated by small $sp^2$ domains comprised of ≤7 aromatic-rings. After 48 h synthesis, the dispersion transition from yellow to red, exhibiting new and red shifted absorbance bands. Furthermore, fluorescence emission is governed by medium-sized $sp^2$ domains (with aromatic ring counts ≤12) and oxygen-containing groups. These oxygen-rich groups within the CQDs, confirmed by FT-IR and XPS, are responsible for the fast photodegradation of organic dyes, with ~90% of methylene blue (MB) being degraded within the first 5 min of light exposure. Our work provides crucial insights about the electrochemical synthesis and overall optical properties of carbon nanostructures, while being effective and reliable toward the degradation of contaminants in water.

*a Instituto de Investigaciones Fisicoquímicas, Teóricas y Aplicadas (INIFTA), Universidad Nacional de La Plata – CONICET, Sucursal 4 Casilla de Correo 16, 1900 La Plata, Argentina. E-mail: fjiban@inifta.unlp.edu.ar*

*b INFIQC-CONICET, Dpto. de Fisicoquímica, Facultad de Ciencias Químicas, Universidad Nacional de Córdoba, Ciudad Universitaria, 5000 Córdoba, Argentina*

*c Resonancias Magnéticas-Centro Atómico Bariloche (CNEA, CONICET), S. C. Bariloche, 8400, Río Negro, Argentina*

*d Instituto de Nanociencia y Nanotecnología, CNEA, CONICET, S. C. Bariloche 8400, Río Negro, Argentina*

*e Instituto Balseiro (UNCuyo, CNEA), Av. Bustillo 9500, S.C. de Bariloche, 8400, Río Negro, Argentina*

*f Instituto de Nanociencia y Materiales de Aragón, CSIC-Universidad de Zaragoza, C/Pedro Cerbuna 12, 50009, Zaragoza, Spain*

*g Laboratorio de Microscopías Avanzadas, Universidad de Zaragoza, Mariano Esquillor s/n, 50018, Zaragoza, Spain*

*h Laboratorio de Química Analítica para Investigación y Desarrollo (QUIANID), Facultad de Ciencias Exactas y Naturales, Universidad Nacional de Cuyo/Instituto Interdisciplinario de Ciencias Básicas (ICB), CONICET UNCUYO, Padre J. Contreras 1300, 5500 Mendoza, Argentina*

*i Dpto. de Física de la Materia Condensada, Universidad de Zaragoza, C/Pedro Cerbuna 12, 50009, Zaragoza, Spain*

*j Instituto de Nanociencia y Nanotecnología CNEA-CONICET, Av. Gral. Paz, 1499, San Martín, Buenos Aires, B1650, Argentina*

† Electronic supplementary information (ESI) available. See DOI: https://doi.org/10.1039/d3tc01774e

## Introduction

Carbon-based nanoparticles (CNPs) are pivotal in multiple applications due to outstanding optoelectronic properties derived from quantum confinement, functional groups, and the type of synthesis performed. As a result, they have become essential in a wide range of applications. CNPs are used as fluorescent probes for medical diagnosis,[1] selective detection of water contaminants,[2,3] as effective photosensitizers in the design of photoanodes,[4] and as efficient photocatalyst for water splitting[5] and photodegradation of contaminants in water[6] among other applications. There is a large number of bottom-up and top-down methods to obtain CNPs that, involves low-molecular-weight molecules in the case of the former and large carbon precursors for the latter.[7] Bottom-up methods include solvothermal, electrochemical, pyrolysis, and microwave-assisted synthesis, among others.[8] Electrochemical-based methods are often preferred due to their low cost, high yield, and ability to provide precise control over reaction parameters, leading to improved selectivity. The proposed mechanism involves electrooxidation, cross-linking, and finally dehydration steps which enables CNPs with high oxygen content.[9] These low-molecular weight precursors involved in the electrochemistry may include, for example, alcohols,[10] citrate,[11] ionic liquids,[12] organic amines,[13] and chiral molecules.[14] Deng et al. first performed electrochemical synthesis directly from ethanol and demonstrated control over the size but not their structure, producing a high yield of amorphous CNPs.[10] Crystalline nature of CQDs significantly alters their physical properties, particularly their optical characteristics. As a result, obtaining crystalline carbon quantum dots (CQDs) through direct electrooxidation of ethanol poses a major opportunity.

Optical properties in CNPs are not as straight forward as in semiconducting quantum dots where the band gap has a direct correlation with size.[15] In fact, they need to be explained by multiple interplaying factors that mainly involve size,[16] edge geometry (armchair or zig-zag),[17] $sp^2/sp^3$ hybridization ratio,[18] number of layers (thickness), and functional groups[19] present within the surface/edge of the carbogenic cores. Kim et al. studied the effect of size and shape on the optical properties of carbon nanostructures.[20] They demonstrated that the absorption wavelength increased (red-shifted) as the size of CNPs increased and observed that circular shapes are comprised of an equal mixture of zig-zag and armchair borders. Recent studies suggest that the optical properties of $sp^2$ carbon networks can be attributed to the collective contribution of various $sp^2$ polycyclic aromatic hydrocarbons (PAHs) present within the structure.[21] These PAHs act as individual absorption/luminescent centers.[21–24] For instance, Urban and co-workers combined three different types of PAHs embedded in a polymer matrix in order to compare their optical properties with CNPs obtained experimentally.[25] In another work, the high crystallinity of PAHs-built CNPs demonstrated outstanding optical properties such as strong absorption bands extending to the visible region and large molar extinction coefficients.[26] Computational tools have been also utilized to describe the optical properties of PAHs that may change with their specific structure or chemistry.[23,27–29] For instance, Peng Chen and co-workers studied how the emission properties of alike-GQDs change with size, edge configuration, functional groups, and addition of heteroatoms.[28] Also, Zhang and co-workers employed time-dependent DFT (TD-DFT) to chemisorb five different types of oxygen-containing groups to the basal plane of graphene dots. They demonstrated that the incorporation and the position of these groups within the nanostructure greatly altered the optical selection rules causing much broader absorption and emission bands as compared to $sp^2$ pristine carbon.[27] Another example is Melia et al. who simulated linear- and hexagonal-shaped PAHs and demonstrated distinct absorption bands based on the number of benzene units and aspect ratio.[23] Despite the significant progress made in the field, characterizing the optical properties of carbon-based nanostructures, by considering the complex interplay between their constituents, remains uncertain.

In this work, we report an electrochemical synthesis of crystalline and ultra-small size CQDs obtained via direct electrooxidation of ethanol to ultimately test them upon photocatalytic degradation of organic dyes. The synthesized product comprises a population of crystalline carbon nanostructures with ≤3.0 nm diameter and 2–3 graphene layers thickness. We investigated the optical properties of the synthesized CQDs by designing and simulating $sp^2$ and $sp^3$ subdomains that are assumed to coexist within the basal plane and the border of these carbon nanostructures. It should be mentioned that simulated oxygen-containing groups were only added at the borders of the PAHs subdomains to prevent the disruption of the $sp^2$ structure. Therefore, different types of $sp^2$ PAHs were simulated in order to search for potential matches with optical bands obtained experimentally. We were able to assign and classify various $sp^2$ subdomains along with oxygen-containing groups potentially present for as-synthesized and 48 h post-synthesis. The presence of hydroxyl- and carbonyl-terminated CQDs are most likely responsible for the photodegradation of methylene blue (MB) via the formation of radical species upon 250–750 nm irradiation lamps. Synthesized CQDs demonstrates outstanding kinetics of photodegradation by removing ~90% of the dye within the first 5 min of irradiation. This work has direct implications for achieving simple and scalable electrochemical synthesis of crystalline nanostructures as well as for further understanding of the optical characteristics of these carbon nanostructures toward optoelectronic applications.

## Experimental methods

### Reagents

All chemical reagents were used as received. Ethanol (99%), Sodium Hydroxide, Methylene Blue were purchased in Sigma-Aldrich Argentina. Aqueous solutions were prepared with ultra-pure water (18.2 mΩ resistivity). Ni foam with thickness = 1.6 mm and porosity = 87% was purchased from MTI Corp. (Richmond, CA, USA). $H_2$ (99.999%) and $CH_4$ (99.999%) were purchased locally from Linde, Argentina.

### Carbon quantum dots (CQDs) synthesis

CQDs were obtained through an electrochemical synthesis formed by Ni foam (anode) and Cu electrodes immersed in 0.1 M NaOH previously dissolved in Ethanol (99% v/v) electrolyte and subjected to 30 V during 60 min. As-synthesized dispersions exhibited basic pH ~ 11.0. For purification of the product, the dispersion was centrifuged at 10 000 rpm in ethanol, filtered with hydrophilic micro-disks, dried off, and redispersed in nanopure water before use. A detailed study regarding ethanol electrooxidation in alkaline media (using Ni foams) along with some insights of CQDs formation mechanisms are included in S.1 and S.2 (ESI†); respectively.

### Characterization

**Structural.** Imaging of nanostructures and structural characterization were performed by High-Resolution Transmission Microscopy (HR-TEM) using an aberration (Objective Lens) corrected Titan³ 60–300 (ThermoFisher) microscope operating at 80 kV at room temperature. Atomic Force Microscopy (AFM) images were obtained using a Nanoscope V (Veeco) microscope in tapping mode. Samples were diluted in absolute ethanol (1:10 000) and drop-cast deposited on mica for AFM analysis.

**Spectroscopy.** UV-vis absorbance spectra were recorded using a PerkinElmer LAMBDA 1050+ UV/Vis/NIR Spectrophotometer. Fluorescence emission and excitation spectra were acquired using a Horiba Fluorolog3 and a standard quartz cuvette. UV-vis and fluorescence samples were diluted in a 1/100 dilution using absolute ethanol. Attenuated Total Reflectance-Fourier Transform-Infrared (ATR-FT-IR) spectra were obtained using a PerkinElmer (Beaconsfield, UK) Spectrum Two instrument with a Universal ATR module. XPS data were collected using a Thermo Fisher Scientific Model K-alpha + with Al Kα radiation at 1486.69 eV (150 W, 10 mA), charge neutralizer and a delay line detector (DLD) consisting of three multi-channel plates. Survey spectra were recorded from −5.0 to 1350 eV at a pass energy of 150 eV (number of sweeps: 2) using an energy step size of 1.0 eV and a dwell time of 100 ms. High resolution spectra for C 1s and O 1s were recorded in the appropriate regions, using an energy step size of 0.1 eV.

**Electrochemical.** Cyclic voltammograms (CVs) were run in a CHI 700D electrochemical station employing a Ni foam and Pt as working and counter electrodes; respectively. The reference voltage is set trough an Ag/AgCl reference electrode submerged in an electrolytic solution of 1.0 M NaOH for the control group and 1.0 M NaOH + 0.5 M ethanol for the test group. For the electrooxidation of ethanol, all the cyclic voltagramms were recorded 10 times in the range of 0–1.6 V.

**Computational.** Calculations were carried out using ORCA quantum Chemistry Package.[30] Ground state geometries of current simulated systems were firstly optimized with DFT using the BP86 functional with a SVP basis set and RIJCOSX approximation using a DEF/JAUX basis set. All dangling bonds were passivated with H atoms. Absorption UV-vis spectra as well as emission spectra were calculated using a TD-DFT method with B3LYP hybrid functional a DEF/J aux basis set without RIJCOSX approximation. Absorbance and fluorescence spectra are calculated using the ASA (Advanced Spectral Analysis) module in ORCA for each PAH.[31]

### Photodegradation of methylene blue

A quartz cuvette containing solutions with different concentrations of MB ($4.7 \times 10^{-6}$, $1.0 \times 10^{-5}$, and $3.0 \times 10^{-5}$ M) and CQDs (0.2, 0.5, and 0.8 mg mL$^{-1}$) were mixed and exposed to UV-vis irradiation. Samples were irradiated with a 100 W Hg discharge lamp (Olympus LH100G) with a wavelength range between 250 to 750 nm. Aqueous solutions of MB and CQDs were stirred during 10 min in the dark previous to the photochemical process, in order to ensure the interaction between dye and the CQDs. The pH of MB was adjusted to 10.5 to match up with the pH of CQDs so greater stability and efficiency can be achieved. Monitoring of MB degradation was followed by UV-vis spectroscopy.

## Results and discussion

### Structural characterization of CQDs

Fig. 1(A) shows a 125 × 125 nm HR-TEM image for CQDs, along with a histogram indicating monodispersed and ~2.8 nm average diameter carbon dots with monomodal narrow Gaussian distribution corresponding to 0.75 nm full-width half-maximum (FWHM) shown below. Fig. 1(B) shows a selected CQD enclosed by a red square which is enlarged in Fig. 1(C) to clearly show its hexagonal symmetry and 0.21 nm lattice fringe corresponding to the reflection of planes of graphene. The observed crystallinity is assessed applying the Fast Fourier Transform (FFT) to the selected CQD as shown in Fig. 1(E). The FFT clearly exhibits a six-fold symmetry pattern and a reciprocal vector of 4.76 nm$^{-1}$ (0.21 nm in real space) corresponding to 01$\bar{1}$0, 10$\bar{1}$0, 1$\bar{1}$00, 0$\bar{1}$10, $\bar{1}$010, $\bar{1}$100 reflections for graphene-like nanostructures. Fig. 1(F) shows inverse FFT obtained by applying a mask to the six diffraction points described before for the same spot. For comparison purposes, a simulated graphene structure is placed on top of the honeycomb lattice. Carbons in the alpha (α) and beta (β) positions of the hexagonal lattice are marked in red and blue; respectively. To further evaluate the crystalline structure of the CQDs, different areas of the sample were randomly examined. We observed, in some cases, that CQDs presented the same ~0.21 nm interlayer spacing but several different crystallographic orientations. Fig. S3 (ESI†) shows a close-up view for selected CQDs showing these features. Depending on the crystal orientation (respect to electron beam direction) some CQDs appeared as mere lattice fringes, this kind of structures only shows a simple FFT pattern composed of points in one direction, as seen in Fig. S4(A)–(C) (ESI†). Additionally, other evaluated areas of the sample exhibited features recently reported as nanodiamonds consistent with the work of Shen et al. who probed that bottom-up synthesis from alcohols could yield cubic carbon phases at low temperatures (423 K).[32] Fig. S4E (ESI†) shows a possible candidate for this type of nanostructure. In general, the vast majority of the evaluated CQDs has shown monodisperse and highly crystalline structures, unlike amorphous carbon phases previously reported employing a similar electrosynthesis.[10]

Fig. 2(A) shows an AFM image of as-synthesized CQDs deposited on atomically flat mica along with a representative cross-section profile marked with a red line shown below. The cross-section indicates CQDs with thickness ranging from 2 to 4 graphene-like layers. Fig. 2(B) shows a histogram from several cross-sections measured from non-agglomerated CQDs depicted in Fig. 1(A). According to the AFM assessment, the sample comprises the majority of non-agglomerated CQDs exhibiting disk-like nanostructures as indicated by the 2–3 layers average height (~1.0 nm). It should be mentioned that higher height profiles of several nanometers were also found and attributed to stacked CQDs.

### Experimental and simulated UV-vis absorbance bands for CQDs

As-synthesized CQDs exhibited crystalline structure and aspect ratio that are reminiscent of PAHs. To explain their optical properties, we designed and simulated the absorbance bands of several sp$^2$ and sp$^3$ subdomains in order to search for a match to the experimentally-obtained bands. Fig. 3(A) and (B) show simulated UV-vis absorbance bands of various PAHs, which fall within the experimental spectra measured for as-synthesized and 48 h post-synthesis; respectively. We chose these two stages based on the distinct color changes of the dispersion experienced over time (see Panel E). At the initial stage of synthesis, the solution looks transparent (Panel E) with no apparent absorbance bands (spectra not shown). As the synthesis continued to culmination at 60 min, the solution turned to a pale yellow (Panel E), and two bands appeared at ∼275 nm and 330 nm as evidenced in Fig. 3(A). Deng et al. also obtained a prominent band close to 280 nm that became more intense with increasing voltage for amorphous carbon dots obtained under similar electrooxidation conditions.[10] In addition, the yellow color and the band at ∼275 nm have been previously observed, in the exact position, for CNPs electrochemically synthesized at room temperature using a graphite anode.[33] This band has been commonly ascribed to the π–π* transition of the aromatic sp$^2$ domains present within the structure and associated with the degree of oxidation[34] and quantum confinement[20,35] of CNPs obtained from different types of syntheses. The main band at 275 nm is assigned to the contribution of coronene (hexagonal configuration of 7 aromatic rings) and naphthalene (2 linear rings). The description for all the simulated PAHs employed in this work is shown in Panel 3F. In addition, secondary bands corresponding to benzo-e-pyrene (∼330 nm) and benzo-ghi-perylene (∼265 nm and ∼374 nm) are also observed. The absorbance bands simulated here for coronene (∼285 nm) falls within PAHs reported by our group[23] as well as UV-vis experiments performed by Patterson[36] and Fetzer.[37] Moreover, pyrene-like molecules observed experimentally at 335,[38] 345,[25] and 360 nm coincide with our simulations.[29] These bands also appeared in similar positions as described by Melia et al. who simulated various linear sp$^2$ (i.e.; anthracene) and hexagonal (i.e.; coronene) carbon nanostructures and determined that absorption bands do not extended beyond ∼400 nm for ∼7 aromatic rings despite their aspect ratio. Small shifts in UV-vis absorbance maximum values may also occur for different concentrations of the same PAH. It has been demonstrated that varying the concentration of a mixture of PAHs (anthracene/pyrene/perylene) perfectly resembles the bands obtained in the as-synthesized carbon nanoparticles.[25]

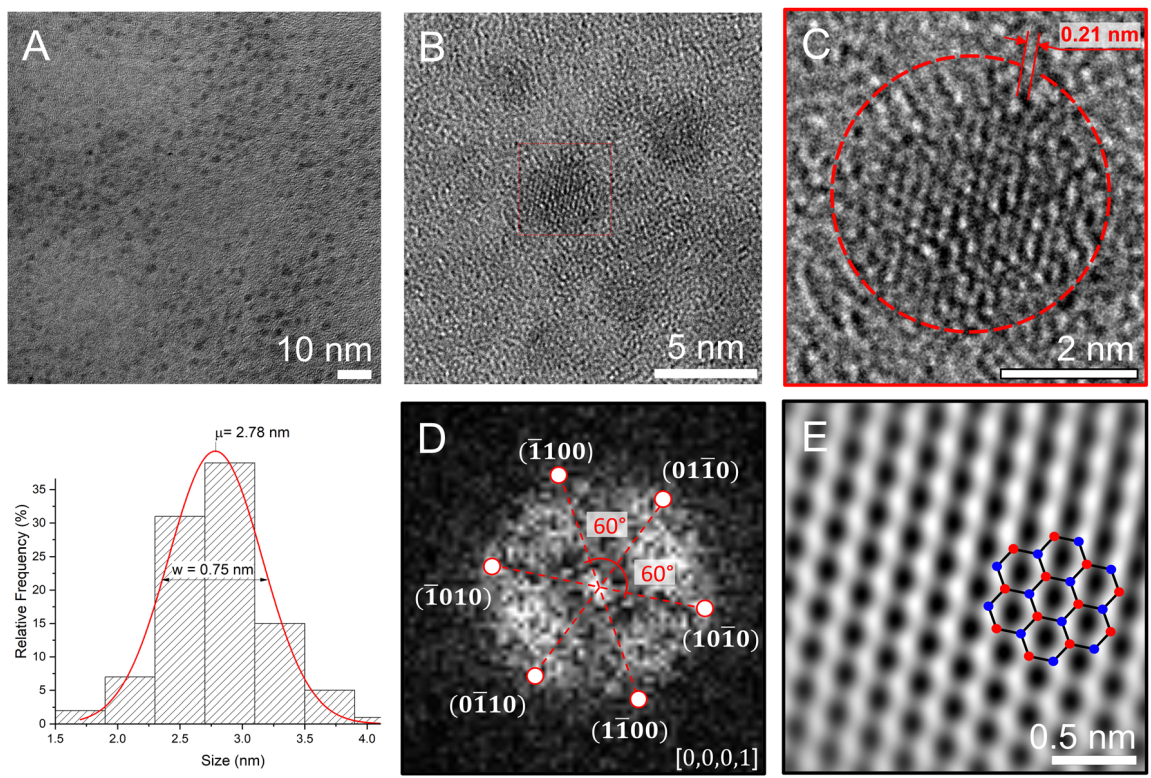

Fig. 1 Zoom-out HR-TEM images of carbon quantum dots (CQDs) (A) along with a histogram indicating ∼2.8 nm average diameter shown below. Selected CQD enclosed by a red box (B) and its magnification showing 0.21 nm lattice fringe (C). Fast Fourier Transform of the selected CQD showing the corresponding six-fold symmetry (D) and inverse FFT (E), digital zoom with the predicted C atom positions.

At 48 h post-synthesis, the dispersion transitioned to red, the entire spectra became flattened and mounted, and the band at ∼331 nm broaden and red shifted toward ∼366 nm as seen in Panel E and Fig. 3(B). The band observed at ∼366 nm can be ascribed to n–π* electronic transition corresponding to the incorporation of oxygen-containing groups (i.e.; carbonyl, epoxide, etc.). This is expected because these CQDs are continuously evolving over time into more complex carbon network of fully mixed sp$^2$ and sp$^3$ subdomains. The particular absorption band observed at 366 nm is attributed to circumbiphenyl PAH, which is circumscribed by hydroxyl groups as noted. There is another broad band that expands from 240 to 265 nm that was assigned to $\pi \to \pi^*$ transition of phenanthrene, chrysene, and pyrene $sp^2$ domains as indicated. It is important to note that those bands within the 240 to 255 nm range, as well as those at 366 nm are absent in the as-synthesized CQDs (Fig. 3(A)). Our simulations indicated that larger PAHs including ovalene, circumcoronene, circumbiphenyl, along with oxygen-containing groups located at the border are the ones responsible for red-shifting the absorbance band. For instance, pristine $sp^2$ pyrene has a prominent band at 336 nm that shifted to 380 nm upon the incorporation of two OH-functional groups at the border (Fig. S5, ESI†). In line with that, Fig. S6 (ESI†) exhibits simulated molecular orbitals of coronene-functionalized carbonyl group. The intermediate band between 325 to 340 nm is filled by n → π* transitions of carbonyl groups modeled by a carbonyl-terminated naphthalene domains (1$H$,3$H$-benzo[$de$]isochromene-1,3-dione).

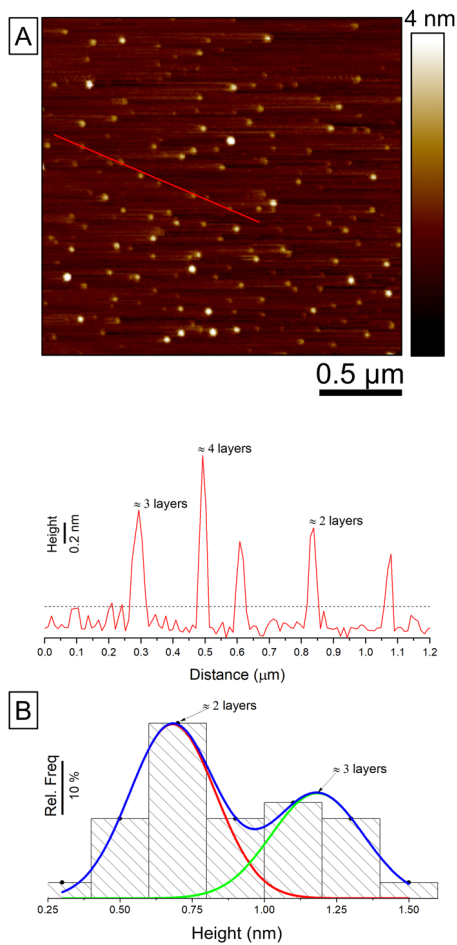

Fig. 2 AFM image (A) along with a representative cross-section profile for as-synthesized CQDs. Histogram of non-agglomerated carbon nanostructures observed in a 2 × 2 μm drop-cast deposited on atomically flat mica (B). Average height indicates carbon structures comprised of 2–3 layers.

### Experimental and simulated emission bands for CQDs

Fig. 3(C) and (D) exhibits the emission spectra of CQDs measured at 48 h post-synthesis at different excitation energies. The emission rate of CQDs depends on the excitation energy.

Our model predicts that different excitation energies will induce emissions from different isolated $sp^2$ subdomains within the carbogenic core. As observed for UV-vis absorbance hydroxyl- and carbonyl-terminated domains red shifted the emission bands. Fig. 3(C) depicts emission spectra for 230 nm excitation, being naphthalene (∼282 nm) and circumbiphenyl (∼340 nm) the dominant features of the spectra. Carbonyl-terminated naphthalene domains and hydroxyl-terminated circumbiphenyl emits in the region of 320 and 360 nm; respectively. Bands located between 340–400 nm emit due to quantum confinement of $sp^2$ subdomains within the carbogenic core.[28] Meanwhile, carbonyl- and hydroxyl-terminated $sp^2$ domains dominate over the extended emission into the visible range of the spectrum. For excitation wavelengths above 360 nm, the emission intensity decreases substantially and becomes dominated by the emission of oxygen-rich functional groups as well as medium size $sp^2$ PAHs subdomains including circumcoronene, circumbiphenyl, and ovalene. Fig. 3(E) and (F) exhibit optical pictures indicating the evolution of color during the electrochemical synthesis and a scheme of a simulated CQD fully comprised of $sp^2$ and $sp^3$ groups; respectively.

### XPS and FT-IR spectroscopy

Fig. 4 compiles XPS and ATR-IR spectra measured for as-prepared CQDs. Fig. 4(A) and (B) exhibits deconvoluted C 1s and O 1s XPS spectra for 48 h CQDs; respectively. Fig. 4(A) indicates the presence of both, $sp^2$ and $sp^3$ peaks seen at 284.4 and 285.0 eV; respectively. It also exhibits two prominent peaks at higher binding energies (BE) observed at 286.8 and 289.0 eV assigned to $sp^3$ hybridization corresponding to epoxide and carbonyl groups; respectively. Fig. 3(B) shows deconvoluted O 1s XPS spectra for CQDs which is dominated by epoxide groups observed at 533.0 eV suggesting a strong oxidation of the surface/edge within the carbon structure. This is consistent with the simulated and experimental bands observed at 48 h after synthesis. A minor peak seen at ∼ 536 eV is assigned to Na Auger corresponding to synthesis byproducts.[39] Assignation of peaks were referenced to Thermo Scientific Avantage Data System for XPS[40] and NIST X-ray Photoelectron Spectroscopy Database.[41] Fig. 4(C) shows ATR-IR bands located at ∼ 880/780, ∼ 1050, ∼ 1110, ∼ 1425–1336, ∼ 1580, ∼ 1778, ∼ 2960–2850, ∼ 3300 (broad) and ∼ 3662 (narrow) corresponding to C–H out of plane bend, C–O stretch of primary alcohols, C–O–C stretching arising from cyclic and alkyl-substituted ethers, C–H asymmetric/symmetric bending, C=C $sp^2$ aromatic ring stretching, C=O stretch, C–H asymmetric/symmetric stretching and bonded/non-bonded O–H hydroxyl groups, respectively. Analysis and peak assignation for ATR-IR were performed according to Coates $et\ al.$[42] These results are consistent with XPS measurements due to several similarities. First, the C 1s XPS records C=C $sp^2$ and $sp^3$ bands, which are confirmed by FT-IR bands located at 1580 cm$^{-1}$ (C=C $sp^2$) and in the region 1420–1050 cm$^{-1}$ were several peaks corresponding to C–H/C–O groups with $sp^3$ hybridized carbon are observed. Furthermore, both XPS and ATR-IR confirms the presence of C–O groups observed at 285.7 eV and 1050 cm$^{-1}$, respectively. Finally, C–O–C groups were also confirmed for bands arising at 286.8 eV (C 1s) and 533.0 eV (O 1s) for XPS and at 1110 cm$^{-1}$ for ATR-IR. Survey XPS spectra for CQDs shows atomic percentage C/O ratio of ∼80/20 (Fig. S7, ESI†). Both techniques XPS and ATR-IR suggest a C=C sp$^2$ hybridized structure fully covered by sp$^3$ hybridized functional groups, indicating that CQDs are comprise of a collection of sp$^2$ domains embedded in a sp$^3$ framework.

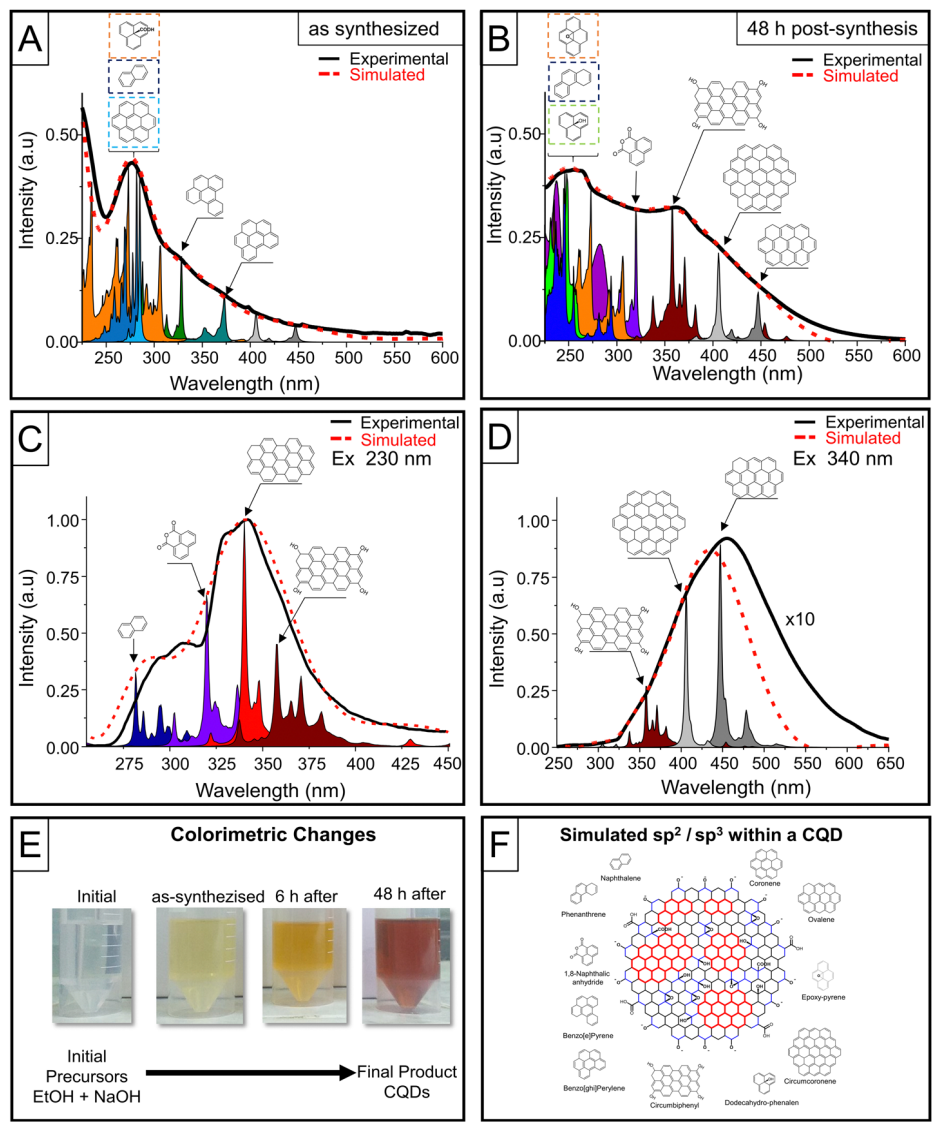

Fig. 3 Experimental and simulated UV-vis absorption for as-synthesized (A) and 48 h post-synthesis (B), averaged contribution of simulated PAHs is depicted in red. Emission spectra for CQDs measured at 230 nm (C) and 340 nm (D) excitation wavelength. Optical pictures showing colorimetric changes at different stages (E). A simulated CQD fully comprised of sp$^2$ aromatic rings (in red) and sp$^3$ (in blue) oxygen-containing groups within the basal plane and border (F).

**Photodegradation of organic dyes**

Fig. 5(A) shows successive representative UV-vis absorbance spectra for optimal concentrations of CQDs (0.8 mg mL$^{-1}$) dispersed in 1.0 × 10$^{-5}$ M MB solution before and after 1, 2, 4, 5, 15, and 30 min exposure to UV-visible lamp (250–750 nm) as indicated. The photodegradation of MB was monitored by looking at changes in intensity of two characteristic bands located at 663 and 613 nm corresponding to MB$^+$ and (MB$^+$)$_2$; respectively. It is expected that the kinetics of photodegradation for most organic dyes follows a pseudo-first order behavior, given by the influence of the initial concentration of the dye, irradiation conditions, nature of the photocatalyst, etc.[6] Based on the Langmuir–Hinshelwood approximation adjusted for adsorption reactions at solid–liquid interfaces, we use the following equation:

$$\ln [C_{MB}/C_{MB}^0] = -k't \qquad (1)$$

where; $C_{MB}$ is the final concentration of MB after a period of time $t$ under irradiation, $C_{MB}^0$ is the initial concentration of MB, $k'$ corresponds to the apparent rate constant of the photocatalytic degradation (min$^{-1}$), and $t$ corresponds to the time (in minutes) of irradiation. Fig. 5(B) shows the degradation of MB at three different starting concentrations (3.0 × 10$^{-5}$, 1.0 × 10$^{-5}$, and 4.7 × 10$^{-6}$ M) later mixed with the same amount of CQDs (0.8 mg mL$^{-1}$). Fig. 5(C) confirms the pseudo-first order reaction as it shows a linear plot of natural log for $C_{MB}/C_{MB}^0$ ratio vs. the initial time under irradiation for up to 2 min of 4.7 × 10$^{-6}$ M and 1.0 × 10$^{-5}$ M MB solutions. The slopes for each curve determine the values of $k'$ and the half life time, $t_{1/2}$, as indicated ($t_{1/2}$ = 0.693/$k'$). Clearly, the photocatalytic activity given by the CQDs is practically equivalent for MB concentrations lower than 1.0 × 10$^{-5}$ M.

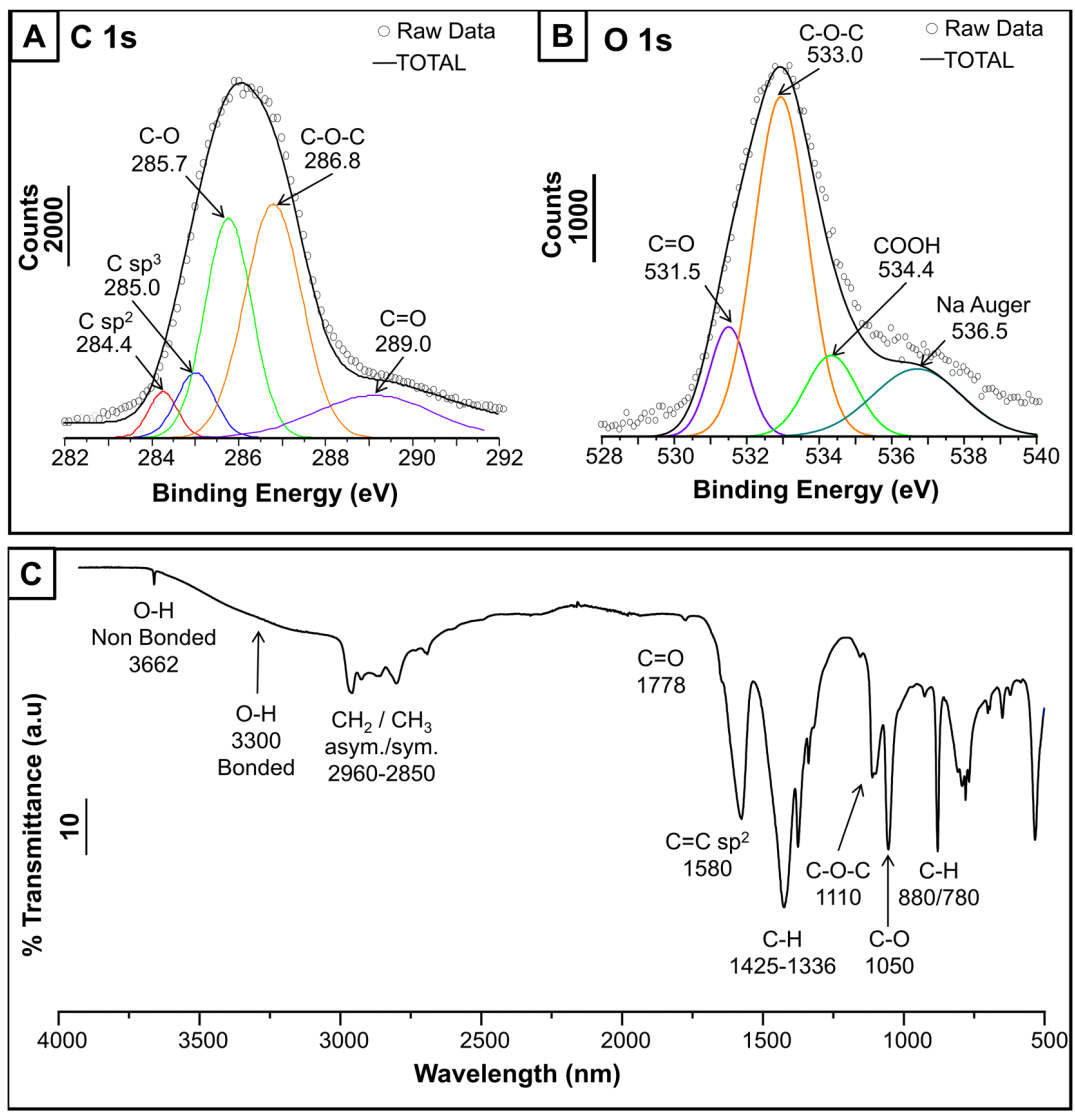

Fig. 4 Deconvoluted C 1s (A), O 1s (B) XPS spectra, and ATR-IR (C) bands for 48 h post-synthesis CQDs.

Fig. 6(A) shows the photodegradation kinetics of 4.7 × 10$^{-6}$ M MB tested at three different concentrations of CQDs as indicated. The general behavior is similar in all cases, with a degradation yield >85% of the dye up to 30 min of irradiation. As expected, at higher CQDs concentrations, the initial rate reaches ∼97% degradation within the first 10 min. This result is quite remarkable from the point of view that most of the MB dye is photodegraded within just 5 to 10 min of UV-vis exposure. Fig. 6(B) compares a plot of $C_t/C_o$ vs. irradiation time for 3 samples: two control experiments with the optimal concentration of CQDs (0.8 mg mL$^{-1}$) performed with and without irradiation on a solution containing MB and irradiating only into a MB solution without photocatalyst. Without irradiation, there were no apparent changes whereas under irradiation the dye initially degraded very slowly (first 5 min) and ultimately reached ∼40% after 30 min of exposure. When the optimal concentration of 0.8 mg mL$^{-1}$ CQDs was mixed with MB, there was a sudden decrease in absorbance intensity down to ∼5% of its initial value within the first 5 min under irradiation. This clearly indicates that CQDs alone are responsible for the fast and efficient photodegradation of the dye. Table S1 (ESI†) compares our work with Literature and highlights some important parameters including the type of carbon dots, type of organic dye, light source and power, and degradation rate and time. A fair analysis would be to compare the performance of our CQDs with those irradiated with lamps of similar wavelength range. In this case, the as-synthesized CQDs outperform others regarding the kinetics of degradation being, in some cases, one order of magnitude faster than other catalysts exposed to similar conditions.[43–46] It is reasonable to expect longer times of degradation when employing low energy lamps such as LED[6] or visible lights[6,43,47,48] due to the low power and longer irradiation wavelengths, respectively. Further studies will explore the irradiation of our catalyst with LED lights. Photodegradation of organic dyes using carbon dots has been an extensively researched area in past years demonstrating fast and steady generation of excitons upon illumination with UV-vis light.[6,49] We believe that the oxygen-containing groups within the CQDs, confirmed by XPS and FT-IR techniques, play a crucial role in producing reactive oxygen radicals upon UV illumination, thus promoting photogenerated electrons from the LUMO to the HOMO band to ultimately transfer electrons and dissolve oxygen in solution as indicated in the scheme next to Fig. 6(B).[6,47]

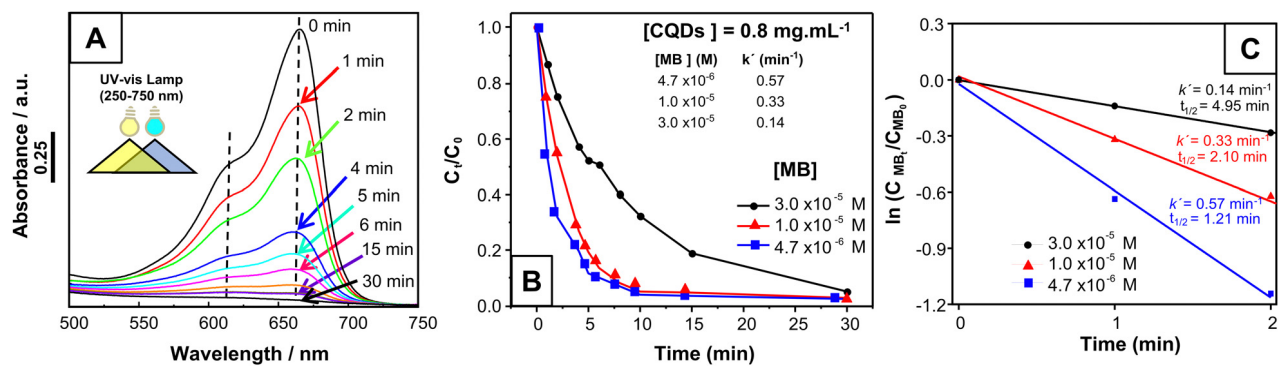

**Fig. 5** Representative UV-vis absorbance spectra for optimal concentration of CQDs (0.8 mg mL$^{-1}$) dispersed in $1.0 \times 10^{-5}$ M MB solution before (0 min) and after 1, 2, 4, 5, 6, 15, and 30 min exposure to 250–750 nm UV-visible lamp as indicated (A). 0.8 mg mL$^{-1}$ CQDs mixed with three different concentrations of MB ($3.0 \times 10^{-5}$, $1.0 \times 10^{-5}$, and $4.7 \times 10^{-6}$ M) to test the rate of degradation (B). Plot of natural log for $C_{MB}/C_{MB}^0$ ratio vs. the initial time under irradiation for up to 2 min indicating linear curves for $4.7 \times 10^{-6}$ M, $1.0 \times 10^{-5}$ M, and $3.0 \times 10^{-5}$ MB solutions (C).

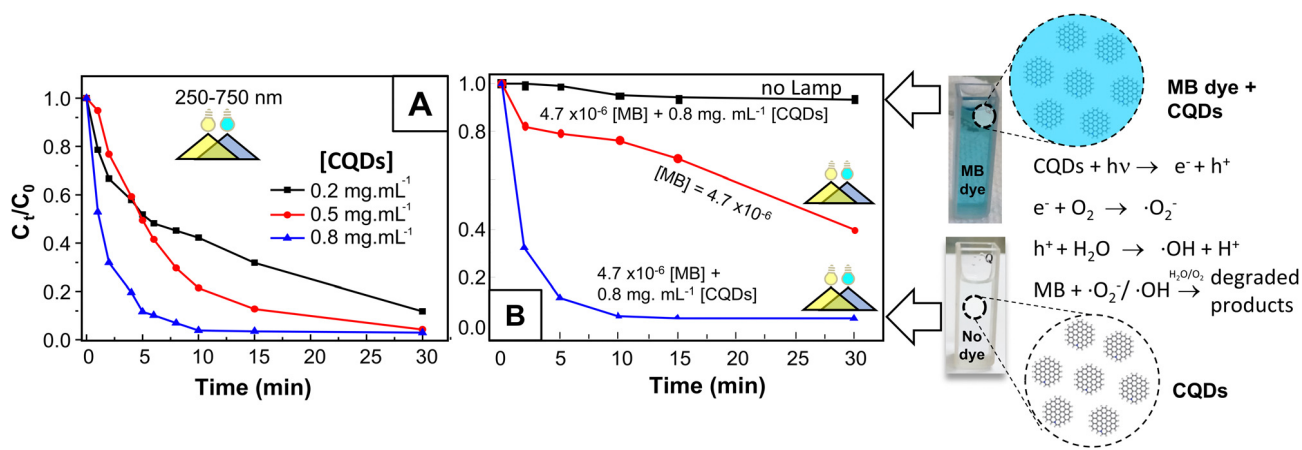

**Fig. 6** Photodegradation of $4.7 \times 10^{-6}$ M MB solution employing three different concentrations of CQDs (as indicated) (A). Plot of $C_t/C_o$ vs. irradiation time for two control experiments performed in dark and by irradiating on the MB solution as well as on the mixture of MB with CQDs at optimal concentrations as indicated (B). The panel next to B shows optical pictures of cuvettes before and after irradiation for 30 min along with the proposed mechanism of free radical generations. The cartoon represents the CQDs before and after the degradation of the dye.

## Conclusions

In this study, we demonstrated a facile and efficient method for the electrosynthesis of graphene-like nanostructures via the oxidation of ethanol on the surface of Ni. In addition, we thoroughly characterized the product using various techniques including aberration corrected HR-TEM, AFM, XPS, UV-vis, Fluorescence, and FT-IR. The obtained product consists of highly crystalline carbon dots with a monodisperse size of approximately 3.0 nm and 2–3 graphene layers thick (0.7–1.0 nm). A considerable number of epoxy- and carbonyl-functional groups populate the as-synthesized CQDs, as indicated by XPS and FTIR measurements. Those functional groups are crucial because they produce an effective π truncation over the sp$^2$ character of the entire structure modifying their optical response from a unique absorbance/emission peak to a sum of coexisting $sp^2$ subdomains present within each CQD. We performed TD-DFT using ORCA quantum chemistry package to calculate absorbance spectra and fluorescence emission of a list of possible candidates with <12 aromatic rings. Our simulations exhibited good agreement with the experimental data and provided a powerful platform to understand their optoelectronic properties. Furthermore, it is demonstrated the potential applications of these CQDs in photo-induced environmental remediation by testing their efficiency in degrading organic dyes. Upon 5 minutes of irradiation, CQDs were able to reduce by 95% the concentration of MB in aqueous solution. Notably, the as-synthesized CQDs exhibited superior kinetics of degradation compared to other catalysts under similar conditions, in some cases, showing one order of magnitude faster degradation. Our study highlights the potential of CQDs for the development of novel organic electronics and their use in environmental remediation applications.

## Author contributions

The manuscript was collaboratively created with contributions from all authors, who have provided their approval for the final version of the manuscript.

## Conflicts of interest

There are no conflicts to declare.

## Acknowledgements

We gratefully acknowledge financial support from PICT (2019–2188) and Proyecto de Incentivos 2019-X-887 from UNLP. SB and FF acknowledge CONICET for their doctoral scholarships. We also acknowledge the financial support of European Commission through Marie Skłodowska-Curie Actions H2020 RISE with the projects MELON (grant no. 872631) and ULTIMATE-I (grant no. 101007825). Authors would like to acknowledge the access of equipment of "Servicio General de Apoyo a la Investigación (SAI), Universidad de Zaragoza" and LAMARX-SNM at Universidad Nacional de Córdoba (UNC), Argentina.

## Notes and references